\documentclass[12pt,preprint]{aastex}

\usepackage{graphics}

\input epsf

\begin{document}

\title{Stellar Populations in Ten Clump-Cluster Galaxies of the Ultra Deep
Field}

\author{Bruce G. Elmegreen \affil{IBM Research Division, T.J. Watson
Research Center, P.O. Box 218, Yorktown Heights, NY 10598, USA,
bge@watson.ibm.com} }
\author{Debra Meloy Elmegreen \affil{Vassar College,
Dept. of Physics \& Astronomy, Box 745, Poughkeepsie, NY 12604;
elmegreen@vassar.edu} }

\begin{abstract}
Color-color diagrams for the clump and interclump emission in 10
clump-cluster galaxies of the Ultra Deep Field are made from
B,V,i, and z images and compared with models to determine
redshifts, star formation histories, and galaxy masses.  These
galaxies are members of a class dominated by 5 to 10 giant clumps,
and having no exponential disk or bulge. The redshifts are found
to be in the range from 1.6 to 3.  The clump emission is typically
40\% of the total galaxy emission and the luminous clump mass is
19\% of the total galaxy mass. The clump colors suggest declining
star formation over the last $\sim0.3$ Gy, while the interclump
emission is redder than the clumps, corresponding to a greater
age. The clump luminous masses are typically $6\times10^8$
M$_\odot$ and their diameters average 1.8 kpc, making their
average density $\sim0.2$ M$_\odot$ pc$^{-3}$. Including the
interclump populations, assumed to begin forming at $z=6$, the
total galaxy luminous masses average $6.5\times10^{10}$ M$_\odot$
and their diameters average $19$ kpc to the $2-\sigma$ noise
level. The expected galaxy rotation speeds average $\sim150$ km
s$^{-1}$ if they are uniformly rotating disks. The ages of the
clumps are longer than their internal dynamical times by a factor
of $\sim8$, so they are stable star clusters, but the clump
densities are only $\sim10$ times the limiting tidal densities, so
they could be deformed by tidal forces. This is consistent with
the observation that some clumps have tails. The clumps could form
by gravitational instabilities in accreting disk gas and then
disperse on a $\sim1$ Gy time scale, building up the interclump
disk emission, or they could be captured as gas-rich dwarf
galaxies, flaring up with star formation at first and then
dispersing.  Support for this second possibility comes from the
high abundance of nearly identical clumps in the UDF field,
smaller than 6 pixels, whose distributions on color-magnitude and
color-color plots are the same as the galaxy clumps studied here.
The distribution of axial ratios for the combined population of
chain and clump-cluster galaxies in the UDF is compared with
models and shown to be consistent with a thick disk geometry.  If
these galaxies evolve into today's disk galaxies, then we are
observing a stage where accretion and star formation are extremely
clumpy and the resulting high velocity dispersions form
thick-disks.  Several clump-clusters have disk densities that are
much larger than in local disks, however, suggesting an alternate
model where they do not survive until today, but get converted
into ellipticals by collisions.
\end{abstract}

\keywords{galaxies: formation --- galaxies: evolution ---
galaxies: high-redshift --- galaxies: irregular}

\section{Introduction}

A high fraction of young galaxies show an irregular clumpy
structure from blue emission that is concentrated into clusters
much larger and more massive than any found locally. Chain
galaxies, for example (Cowie \& Hu 1995; van den Bergh et al.
1996; Moustakas et al. 2004) are dominated by several giant blue
clumps strung along a line; there are no exponential light
profiles or central red bulges as in spiral galaxies today
(Elmegreen,  Elmegreen, \& Sheets 2004, hereafter Paper I).
Dalcanton \& Schectman (1996) suggested that chains are edge-on
low surface brightness galaxies. O'Neil et al. (2000) considered
high-redshift disk galaxies in general, many of which have bulges
and exponential profiles, and concluded from the distribution of
axial ratios that chains are edge-on disks. Reshetnikov, Dettmar,
\& Combes (2003) showed that the axial ratios of 34 edge-on
galaxies in the Hubble Deep Field are larger than for local
galaxies, implying thicker disks, and noted that half of their
sample, including some face-on galaxies, have non-exponential
profiles like the chains. Edge-on local galaxies resemble chain
galaxies too when seen in projection at ultraviolet wavelengths,
although the local star forming regions are much smaller than in
chain galaxies (Smith et al. 2001).

Elmegreen, Elmegreen \& Hirst (2004a, hereafter Paper II) studied
a large sample of chain and non-exponential clumpy galaxies in a
deep Advanced Camera for Surveys image and confirmed that they are
members of a new class of young galaxies with thick,
non-exponential disks dominated by $\sim10^9$ M$_\odot$ blue
clusters. They have no regular bulge structures, like other disk
galaxies at the same redshifts (Paper I), but sometimes have
irregular or offset bars with a bulge-like size (Elmegreen,
Elmegreen, \& Hirst 2004b, hereafter Paper III). An objective name
for this type was suggested to be a ``clump cluster,'' because
without kinematic data it was not known whether each is a
conglomeration of pre-existing small galaxies, like a tiny
Stephan's Quintet, or a single coherent galaxy. A lack of clear
rotation in some linear galaxies at high redshift suggests they
could be conglomerates (Bunker et al. 2000; Erb et al. 2004).
Chain galaxies also lack rotation in the models by Taniguchi \&
Shioya (2001), where the objects are shocked filaments between
expanding galactic-wind bubbles, and by Iono, Yun, \& Mihos
(2004), where they are interaction-induced bars with gas inflow.
If they are in fact coherent objects, then ``clumpy galaxies''
might be a better term. Here we refer to them as clump-clusters or
clump-cluster galaxies, depending on the context.

Conselice et al. (2004) noticed a similar dominance of clumps in
many young disk galaxies, including galaxies with bulges and
exponential-like profiles, and called them ``luminous diffuse
objects.'' Roche et al. (1997) found that faint and large blue
galaxies with low surface brightness tend to have spiral or
chain-like morphologies. Paper I showed that at I band, chains
dominate all edge-on galaxy types by a factor of $\sim2$ beyond
24th magnitude. Conselice et al. (2004) showed that luminous
diffuse objects dominate star-forming galaxies at redshifts
$z>1.5$.  Conselice, Blackburne, \& Papovich (2004) found that
peculiar galaxies dominate at $z>2$. Consistent with this, there
are very few nearby chain galaxies having I$<22$ mag (Abraham et
al. 1996).

To study further the formation of stars and disks in young
galaxies, we surveyed the Ultra Deep Field (UDF; Beckwith et al.
2004) for more examples of chains and clump-clusters. We
identified 178 clump clusters with major axes exceeding 10 pixels,
and measured their integrated colors, magnitudes, and surface
brightnesses as well as the properties of the individual clumps
and interclump regions.  The lack of exponential disks was
confirmed with contour plots and radial profiles. Visual
inspection is preferred for clumpy galaxies because the SExtractor
program sometimes mistakes them for multiple sources (Paper I).
Here we discuss 10 examples of this class. A full catalog will be
presented in Elmegreen et al. (2005). The UDF is processed for a
resolution of 0.03 arcsec per pixel; at a redshift between 1 and
3, 0.03 arcsec corresponds to a spatial scale between 220 pc and
230 pc in a $\Lambda$CDM model with $\Omega_\Lambda=0.73$,
$\Omega_M=0.27$, and $H_0=71$ km s$^{-1}$ Mpc$^{-1}$ (Spergel et
al. 2003).

Galaxy evolution models of the colors and magnitudes of distant
galaxies have been used to estimate the redshifts, ages,
extinctions, and stellar masses. Studies of galaxies in the Hubble
Deep Fields North (Papovich, Dickinson \& Ferguson 2001; Dickinson
et al. 2003) and South (Rudnick et al. 2003), the Chandra Deep
Field South (Idzi et al. 2004), and in fields observed from the
ground (Shapley et al. 2001) show a general trend of decreasing
star formation rates and increasing stellar mass over time to the
present (see review in Ferguson, Dickinson, \& Williams 2000). We
are interested in the ages and masses of the clump and interclump
regions in our sample and use evolution models like these to
compare with the observations. We would like to determine if the
clumps formed in initially smooth disks as a result of spontaneous
gaseous instabilities (Noguchi 1999; Immeli et al. 2004a,b), or if
they formed by some other mechanism, such as compression-triggered
star formation in giant, hierarchically-merging clouds (Brook et
al. 2004). The clumps could also be separate galaxies combined
during minor mergers (Walker, Mihos, \& Hernquist 1996; Abadi et
al. 2003). The instability model might apply if the gas accretion
was relatively smooth, as in galaxy formation models by Murali et
al. (2002), Westera et al. (2002),
Sommer-Larsen, G\"otz, \& Portinari (2003),
and Keres et al. (2004). Smooth
gas accretion on the scale of the disk could occur in the
hierarchical model if the gaseous parts of the accreting galaxies
are stripped from the stars at large distances from the disk and
then settle hydrodynamically.

Conselice (2004) suggested that the early build-up phase of galaxy
formation, corresponding to $z>3$, is dominated by major mergers,
the intermediate phase at $1<z<2$ is dominated by hydrodynamic gas
accretion, and the late phase ($z<1$) is dominated by minor
mergers.  He noted that Luminous Diffuse Objects, which are
morphologically similar to clump clusters, have bright outer
regions with low central light concentrations, and so might be
examples of the intermediate phase.  The signature for this may be
present in the age and mass distributions of the clump and
interclump regions, studied here.

\section{Data Analysis and Results}
\label{sect:analy}

The UDF images used for this study were observed and processed by
Beckwith et al. (2004) and are available on the STScI archive. The
UDF consists of images in 4 filters: F435W (hereafter B$_{435}$
band; 134880 s exposure), F606W (V$_{606}$ band; 135320 s), F775W
(i$_{775}$ band; 347110 s), and F850LP (z$_{850}$ band; 346620 s).
The magnitude zero points are posted on the archival data sites:
25.673 (B), 26.486 (V), 25.654 (i), 24.862 (z). The images were
mosaiced and drizzled by the HST team to produce final images of
$10500\times10500$ pixels with a scale of 0.03 arcsec per px.
Beckwith et al. used the program SExtractor to identify some
10,040 sources, so we use their nomenclature for the sources
we found independently by visual examination.

\begin{figure}
\epsscale{0.7} \plotone{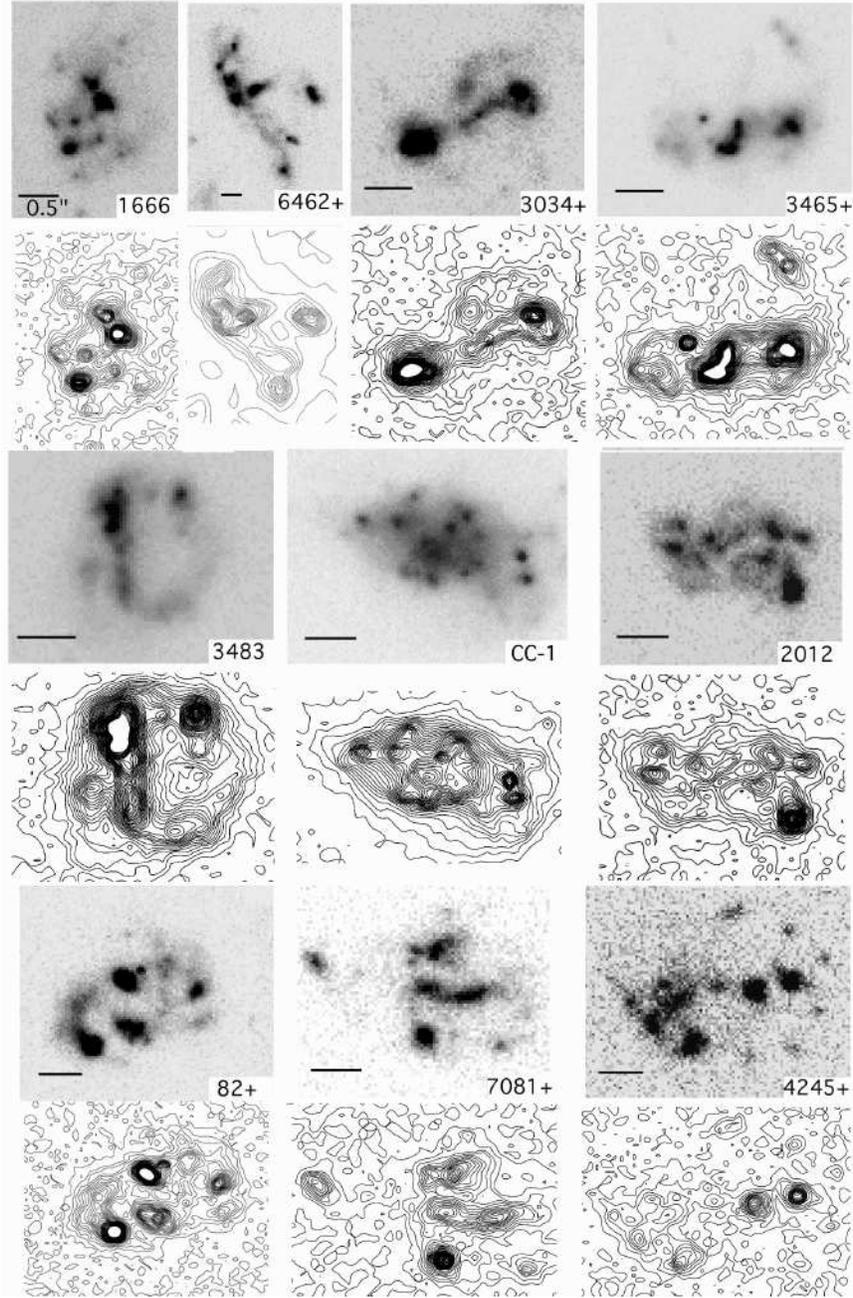} \caption{Grayscale i-band images
and contour plots for 10 clump-cluster galaxies. The bar indicates
$0.5^{\prime\prime}$; galaxy UDF names are noted. The contours are linear in
steps of 1 $\sigma$ sky noise, with the outermost contour at 26.8
mag arcsec$^{-2}$. }\label{fig:gray}\end{figure}

Figure \ref{fig:gray} shows 10 of the most spectacular clump
clusters seen at i$_{775}$-band. They were chosen because of their
large angular size, good resolution into clump and interclump
regions, and wide range of feature types representing most
galaxies in this class. Their positions and integrated properties
are in Table \ref{tab:prop}.  Some have multiple UDF numbers
because SExtractor identified the individual large clumps as
separate sources. Contour plots of intensity are drawn with the
lowest contour 1$\sigma$ above sky noise, corresponding to a
surface brightness of 26.8 mag arcsec$^{-2}$ at i$_{775}$-band.
The lower left-hand bar in each panel represents 0.5 arcsec. The
numbers in the lower right are the UDF catalog numbers.

Objects UDF 3483 and UDF 82+99+103 are ring-like galaxies, while
UDF 7081+7136+7170, UDF 4245+4450+4466+4501+4595, UDF 3034+3129,
UDF 3465+3597, and UDF 3483 have linear central sections
resembling bars (Paper III). UDF 4245+ also has several small
clumps distributed around the periphery, while galaxy CC-1 has
many small clumps throughout the body. UDF 2012 apparently has a
dust lane running through the middle.

The integrated galaxy magnitude was measured in a rectangular box
with edges defined by the i$_{775}$ isophotal contours $2\sigma$
above the sky noise for each filter (corresponding to a surface
brightness of 26.0 mag arcsec$^{-2}$). We also did photometry on
each prominent clump and several interclump regions, giving
surface brightnesses and colors.  Clump photometry was done using
$imstat$ in IRAF by measuring the total flux inside a box that
surrounds the clump and then subtracting the adjacent background
in the same size box. The background was based on several flux
readings in the immediate neighborhood of the clump.

Bright clumps have surface brightnesses that are $\sim1.5$ mag
arcsec$^{-2}$ brighter than the interclump regions. On average,
$\sim40$\% of the i-band luminosity is in the clumps, as indicated
by the clump luminosity fractions, $f_{\rm lum}$, in Table 1. This
luminosity fraction corresponds to a different rest-frame
wavelength in each case because the galaxy redshifts differ, so it
is only meant to be a indicator of the highly clumped morphology
viewed at i-band. The two cases with small $f_{\rm lum}$, UDF
4245+ and CC-1, have many more clumps, and each is much smaller
than in the other objects. Most clump colors within a given object
are similar to each other, although some apparently dusty
interclump regions are much redder than others, such as the dust
lane in galaxy UDF 2012.

\begin{figure}\epsscale{0.8}
\plotone{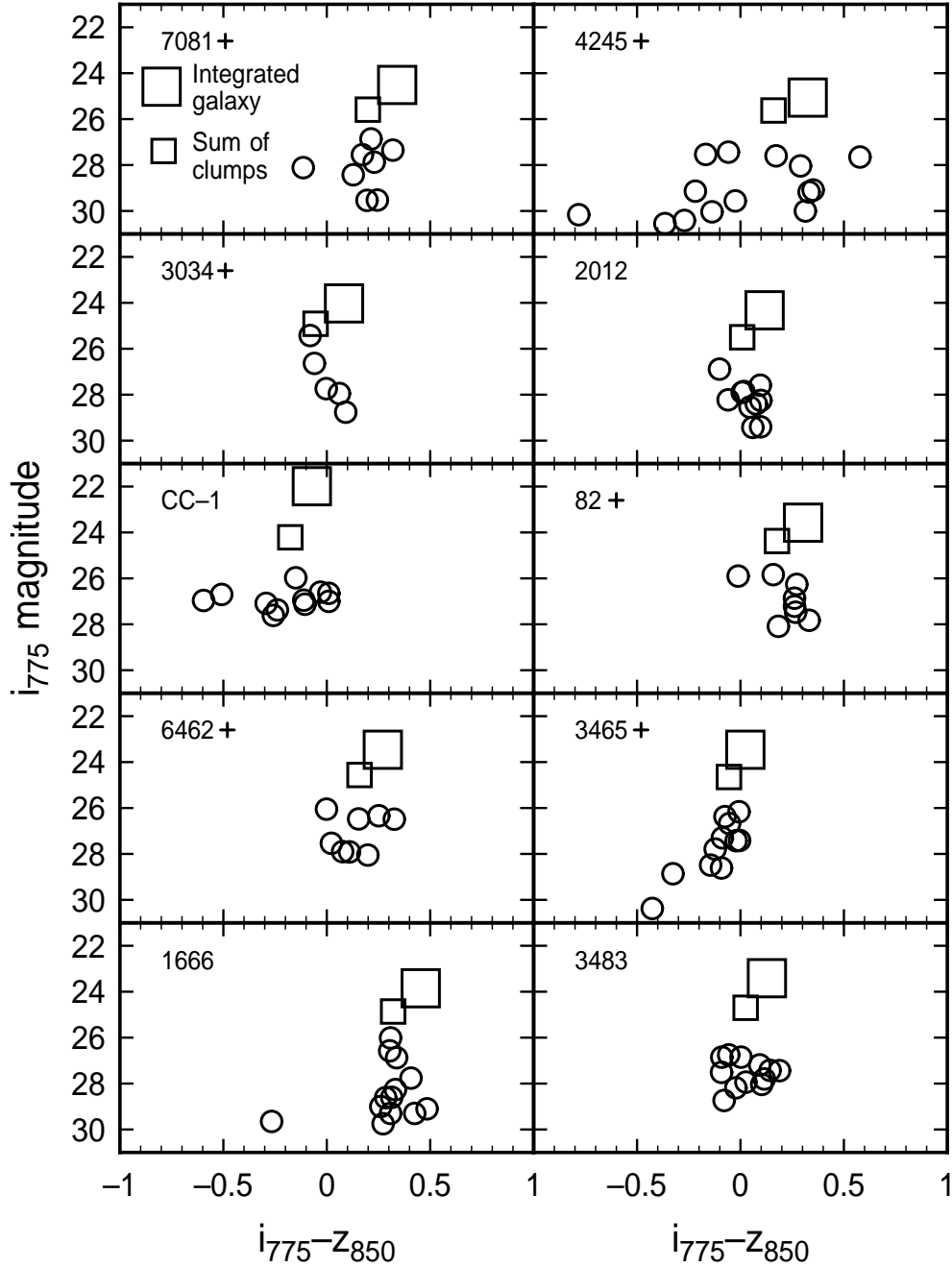}\caption{Color-magnitude distribution for clumps (circles), for the
sum of all clumps (small squares) and for the whole galaxy (big squares).
Abbreviated galaxy names are indicated. }\label{fig:cm}\end{figure}

A color-magnitude diagram for i$_{775}$ versus i$_{775}$-z$_{850}$
is shown in Figure \ref{fig:cm} for all of the big clumps in
Figure \ref{fig:gray} (circles) and for the sum of these clumps
(small squares) and the integrated galaxies (large squares). The
i$_{775}$ magnitudes of the clumps range from 25.4 to 30.5. The
colors lie mostly in a narrow range for each galaxy but they vary
from galaxy to galaxy with the redshift (see below).

\begin{figure}\epsscale{0.8}\plotone{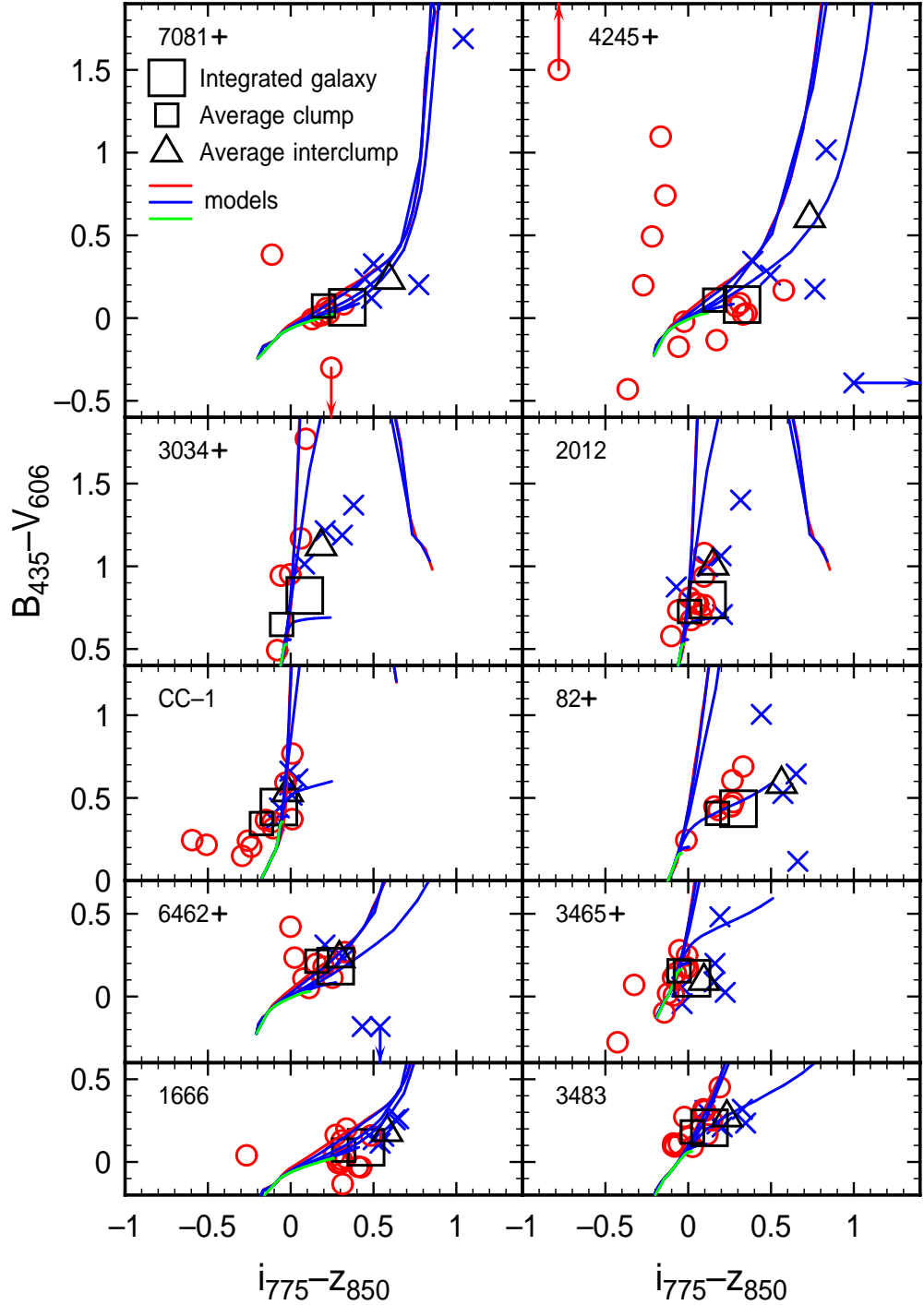}\caption{Color-color
distribution of clumps (circles), interclump regions (crosses), and whole
galaxies (squares). The average colors for the clumps (small squares)
and interclump regions (triangles)
are also shown. Curves are theoretical models that best match the fitted redshifts
of the galaxies to the nearest 0.25. In general, the lower curves
(bluer) correspond to the largest star formation decay times, which
are assumed to be infinite (green curve for the electronic version of this
publication), $10^7$ yr, $3\times10^7$ yr, $10^8$ yr,
$3\times10^8$ yr, and $10^9$ yr (red curve). The best-fit
models are determined simultaneously from both this color-color plot and
the one in the next figure. Observed points that are far off the
range of scales are indicated by arrows; they were not included in the
model fits.  }\label{fig:bvizoverlaydust_best}\end{figure}

\begin{figure}\epsscale{0.8}\plotone{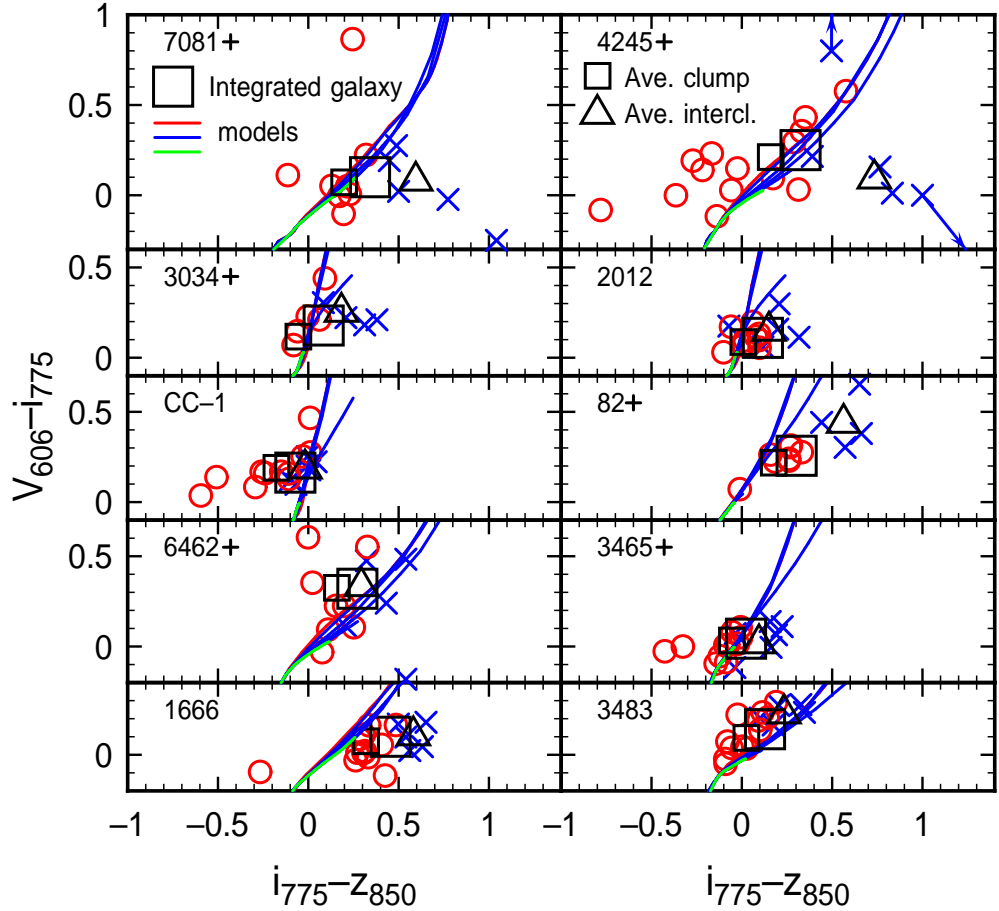}\caption{Color-color
distribution of clumps (circles), interclump regions (crosses), whole
galaxies (squares) and averages (small squares, triangles) are shown
as in Figure 3, along with theoretical tracks that are closest in
redshift to the fitted values. }\label{fig:viizoverlaydust_best}\end{figure}

Color-color diagrams for each clump-cluster galaxy are in Figures
\ref{fig:bvizoverlaydust_best} and \ref{fig:viizoverlaydust_best}.
The circles are the clumps, the crosses are the interclump media
at various positions, the big squares are the integrated galaxies,
the small squares are average colors of the clumps (weighted by
clump luminosity), and the small triangles are the average
weighted colors of the measured interclump emissions. Superposed
curves are best-fit models that will be described in Section
\ref{sect:models}. The reddest B$_{435}$-V$_{606}$ colors are in
UDF 3034+3129 and UDF 2012.  In UDF 3034+3129, the central clump
in the image has B$_{435}$-V$_{606}=1.8$ mag, in UDF 2012, most of
the red interclumps at B-V$>1$ mag are in the dust lane. Also very
red (B$_{435}$-V$_{606}=1.1$ mag) is the large western clump in
UDF 4245+, which has a similarly red interclump region. The two
bluest clumps in B$_{435}$-V$_{606}$ are in UDF 7081+ (the
southernmost round clump) and UDF 2012 (the central clump just
above the dust lane). The bluest clump in $i_{775}-z_{850}$ is in
CC-1, and is a faint clump in the southeast. Figures
\ref{fig:bvizoverlaydust_best} and \ref{fig:viizoverlaydust_best}
show that the interclump regions are systematically redder than
the clumps.

\section{Comparison with models}
\label{sect:models}

\subsection{Model Components}

Photometric redshifts for the 10 clump-cluster galaxies and the
ages and luminous masses of the clumps and interclump regions were
determined by comparison with stellar population models. The
luminous mass is defined to be the mass that fits the observed
emission in the bandpasses we have available. This mass does not
include possible contributions from older populations of stars
that do not emit significantly in the observed bandpasses and are
not modelled explicitly in the star formation history.

The models used the GALAXEV tables of spectra in Bruzual \&
Charlot (2003) for the Padova1994 evolutionary tracks with a
Chabrier (2003) IMF (extending from 0.1M$_\odot$ to 100 M$_\odot$
with a flattening below $\sim0.5$ M$_\odot$).  These tables give
integrated population spectra at 1221 wavelengths from 91 \AA\ to
160 $\mu$m and at 221 times from $1.25\times10^5$ years to
$2\times10^{10}$ years.  We use the low-resolution version. The
spectra were integrated over time with two basic time dependencies
for star formation, a constant rate, and an exponentially
decreasing rate with timescales of $10^7$, $3\times10^7$, $10^8$,
$3\times10^8$, and $10^9$ years.  A choice of six metallicities is
available in GALAXEV. Shapley et al. (2004) found that the
metallicities for 7 disk galaxies at the same redshifts as those
found here ($z\sim2-2.5$) are approximately solar ($Z=0.02$). The
metallicity of the Milky Way thick disk, which may have formed at
about this epoch (Dalcanton \& Bernstein 2002; Brook et al. 2004)
peaks at $[{\rm Fe/H}]\sim=-0.6$ ($Z=0.005$). We use $Z=0.008$ in
what follows; different metallicities did not change the results
significantly.

\begin{figure}\epsscale{0.8} \plotone{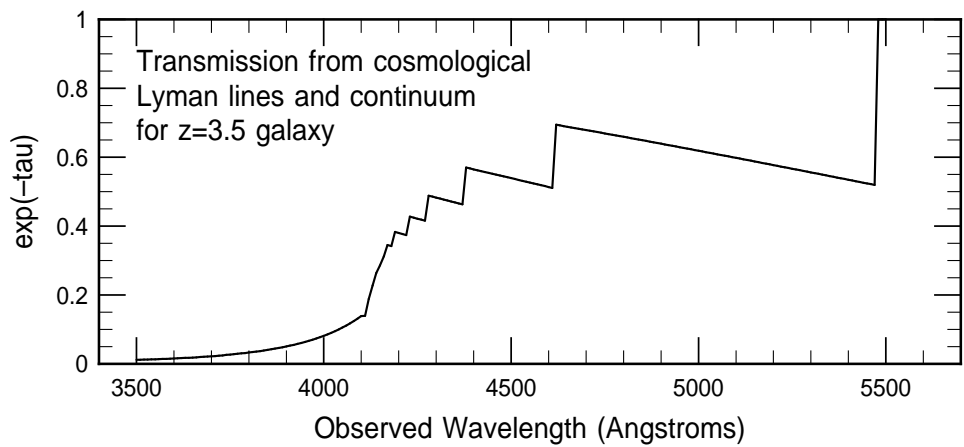}\caption{Transmission from foreground
cosmological hydrogen clouds for a galaxy at redshift $z=3.5$
calculated using the method of Madau
(1995).}\label{fig:madau}
\end{figure}

\begin{figure}\epsscale{0.8}\plotone{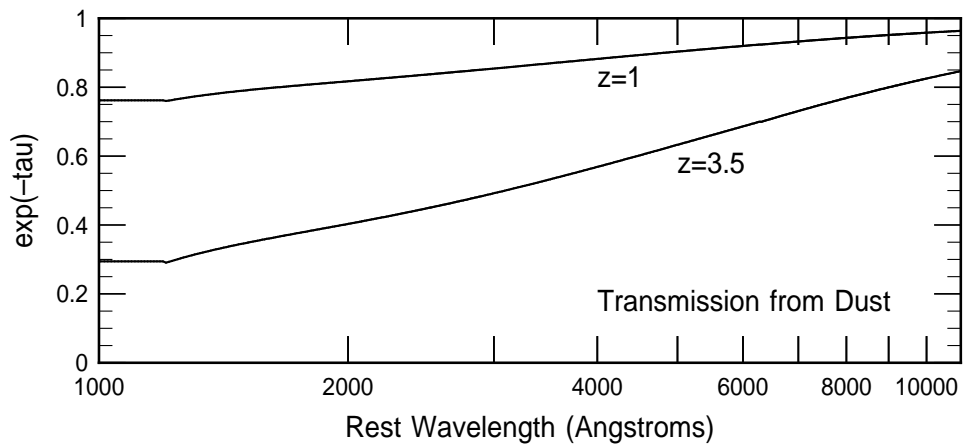}\caption{Transmission from dust
inside the model galaxies using the Calzetti et al. (2000) and
Leitherer et al. (2002) extinction curves and the
intrinsic $A_V$ extinction in Rowan-Robinson (2003). Dust produces
less spectral structure and color variation than cosmological hydrogen
absorption, so the redshifts derived here are not
sensitive to internal extinction.}\label{fig:dust}\end{figure}

Absorption from cosmological hydrogen lines was included according
to the prescription in Madau (1995), taking Lyman series lines up
to order 20 and including continuum absorption.  Figure
\ref{fig:madau} shows the resulting transmission for a source at
$z=3.5$ for comparison with Figure 3 in Madau (1995). Dust
absorption was included using the wavelength dependence in
Calzetti et al. (2000) with the short-wavelength modification in
Leitherer et al. (2002) and the redshift dependence for galaxy
intrinsic $A_V$ extinction in Rowan-Robinson (2003). The resulting
rest-frame dust transmissions for galaxies at $z=1$ and $z=3.5$
are shown in Figure \ref{fig:dust}.  Other estimates of extinction
in $z\sim2$ disk galaxies are 2 or 3 times higher than these
(e.g., Adelberger \& Steidel 2000; Papovich, Dickinson, \&
Ferguson 2001), so two other sets of models with twice and four
times the Rowan-Robinson extinction were fit to our data for
comparison, as discussed below.  In general, higher extinction
lowers the age and mass-to-luminosity ratio as the intrinsic
stellar populations become bluer and brighter, but the masses may
either increase or decrease depending on the relative size of the
changes in $M/L$ and $10^{0.4A}$ for extinction $A$; the redshift
solutions do not change noticeably with extinction over this
range.

\subsection{Model color evolution}

The time-integrated spectra, corrected for cosmological hydrogen
absorption and intrinsic galactic dust, were convolved with the
four throughput functions for B$_{435}$, V$_{606}$, i$_{775}$, and
z$_{850}$ filters, as determined from the IRAF task SYNPHOT
(kindly run for us by Jennifer Mack at STScI). Equation 8 in the
postscript file supplement of the GALAXEV download was used in the
AB magnitudes convention. Because this AB version was not written
explicitly for intensity as function of wavelength, we give it
here:
\begin{equation}
m=-2.5\log\left[{{\int_{-\infty}^{\infty} d\lambda
\lambda{{L_\lambda\left[\lambda\left(1+z\right)^{-1}\right]} \over
{\left(1+z\right)c4\pi d_L^2}} F(\lambda)} \over
{\int_{-\infty}^{\infty} {{d\lambda }\over{ \lambda }}
F(\lambda)}}\right]-48.60.\label{eq:bc}
\end{equation}
To get the constants of proportionality, the luminosity $L$ in
GALAXEV, which is tabulated in units of L$_\odot$ $\AA^{-1}$ for
each solar mass of stars, was converted to erg s$^{-1}$ $\AA^{-1}$
using $L_\odot=3.826\times10^{33}$ ergs s$^{-1}$; $c$ is the speed
of light in cm s$^{-1}$, $d_L$ is the luminosity distance in cm.
The filter function, $F(\lambda)$, is dimensionless, as is
$d\lambda/\lambda$ in the denominator. The combination $d\lambda
L_\lambda$ has dimensions of ergs when $\lambda$ is in $\AA$, but
the other $\lambda$ in the numerator has to be converted to cm to
give $\lambda/c=1/\left(3\times10^{18}\right)$ s$^{1}$;
$\lambda$ is the observed wavelength.

\begin{figure}\epsscale{0.8}\plotone{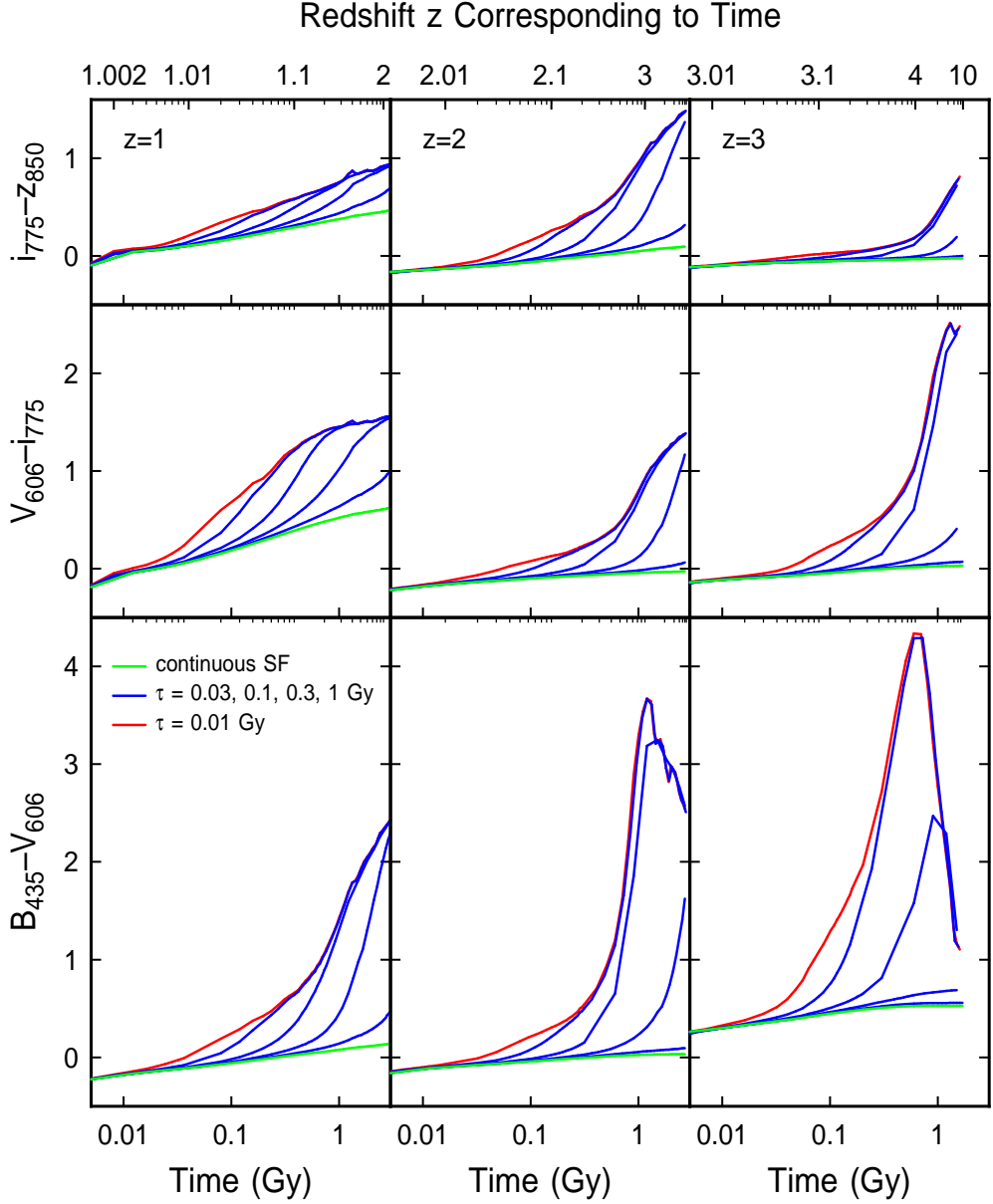}\caption{Model colors
versus population age for three galaxy redshifts, $z=1$, 2, and 3,
and six models of the star formation history, as in Figure 3.
Redshifts corresponding to the population ages for given galaxy
redshifts are on the top axis. The ticmarks for these redshifts
are not the same interval throughout, but they should still be
self-evident; for example, on the top left, the ticmarks are at
1.002, 1.003, 1.004, etc., to 1.01, and then 1.01, 1.02, 1.03,
etc. to 1.1, and then 1.1, 1.2, 1.3, etc. to 2.
}\label{fig:ct}\end{figure}

Models of color versus time are in Figure \ref{fig:ct} for three
values of redshift and for the six star formation histories. The
lower curve in each panel (bluest color) is for continuous star
formation and the upper curve (reddest color) is for the most
rapidly decaying star formation, with an exponential decay time of
10 My. Intermediate curves span the range of exponential decay
times, $\tau$, indicated in the figures. The colors rise with time
rapidly, especially at high redshift, because of the depletion of
uv light in the absence of current star formation. The Lyman
absorption jump shown in Figure \ref{fig:madau} reddens the
B$_{435}$-V$_{606}$ values for all of the curves when $z>3.5$ (the
drop-out effect -- Steidel et al. 1996). The redshifts of star
formation corresponding to the times on the abscissae are
indicated on the top axes. Analogous models without intrinsic dust
extinction differed only a little in color, at most $\sim0.1$ mag
in B$_{435}$-V$_{606}$.

Model color-color plots are shown in Figures \ref{fig:bviz} and
\ref{fig:viiz}.  Stellar age in the rest frame of the galaxy
increases along the curves from 0 at the bottom left (bluest
colors) to whatever age corresponds to $z\sim8$ at the observed
redshift (5.3 Gy, 2.7 Gy, and 1.5 Gy at $z=1$, 2, and 3,
respectively). The different star formation histories have
separate curves. Solid curves have dust extinction from
Rowan-Robinson (2003) and dashed curves have 4 times this
extinction.  Generally, the more continuous the star formation,
the closer the curve stays to the lower left end, even at large
times. The most extreme red values occur for the most rapid decay,
where the exponential time is $\tau=10$ My.

\subsection{Photometric redshifts}

The observed colors of the clump and interclump emissions
correspond to the lower left parts of the theoretical color-color
curves. Figures \ref{fig:bvizoverlaydust_best} and
\ref{fig:viizoverlaydust_best} show the data and models together,
using the redshift models that best match the colors. To find
these redshifts, we made models with the six star formation
histories mentioned above and with redshifts ranging from $z=0$ to
6.75 in intervals of $z=0.25$. Model colors were tabulated for a
wide range of times within each history type. The rms deviations
between the average clump positions and the models on both
color-color plots were then determined for each clump-cluster
galaxy, and the model with the minimum rms was chosen. The highly
deviant colors shown by arrows in the figures were not used for
this. Thus the photometric redshifts are based only on the clump
colors, not the average galaxy colors or the interclump emission.
We do this because clumps are more homogeneous than whole
galaxies. The fit to the redshift also gives the preferred star
formation decay time, $\tau$, and the average age of the clumps in
each object.

\begin{figure}\epsscale{0.8}\plotone{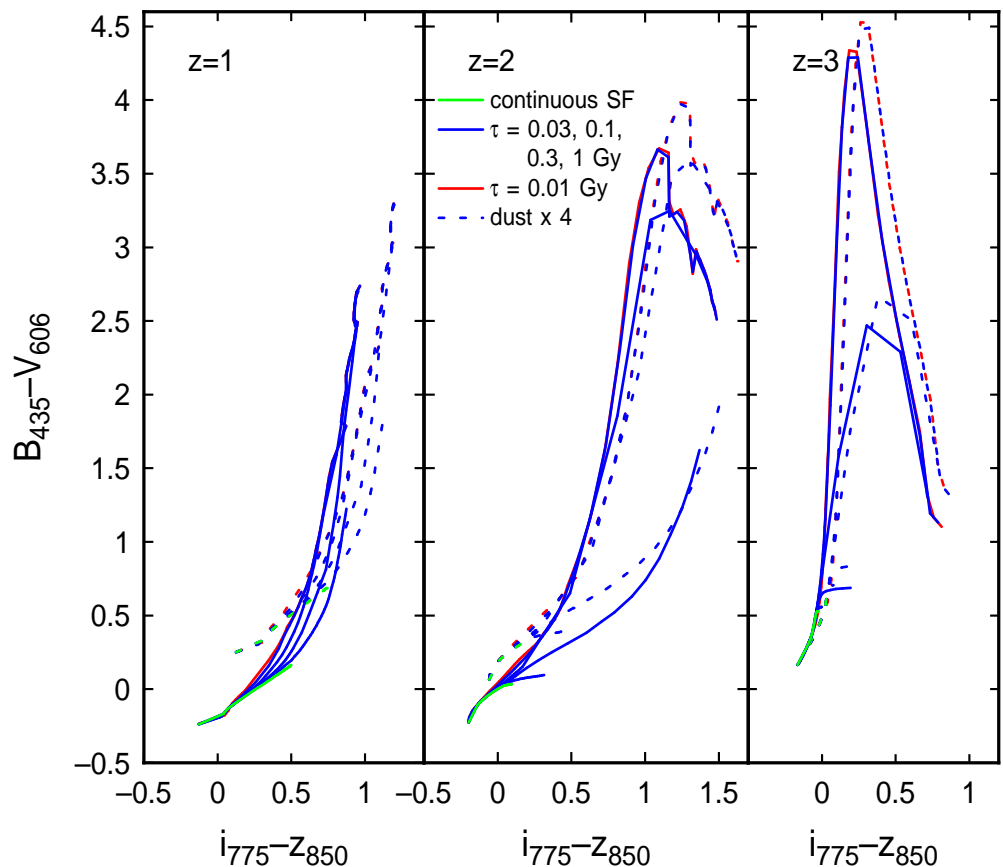}\caption{Population
models on a color-color plot where time increases along each curve from the
lower left to the upper right. The same six star formation histories
as in the previous figures are used, with the same color coding. The solid
lines are for internal extinction following the Rowan-Robinson (2003)
result, and the dashed lines are for internal extinctions that are
4 times higher. }\label{fig:bviz}\end{figure}

\begin{figure}\epsscale{0.8}\plotone{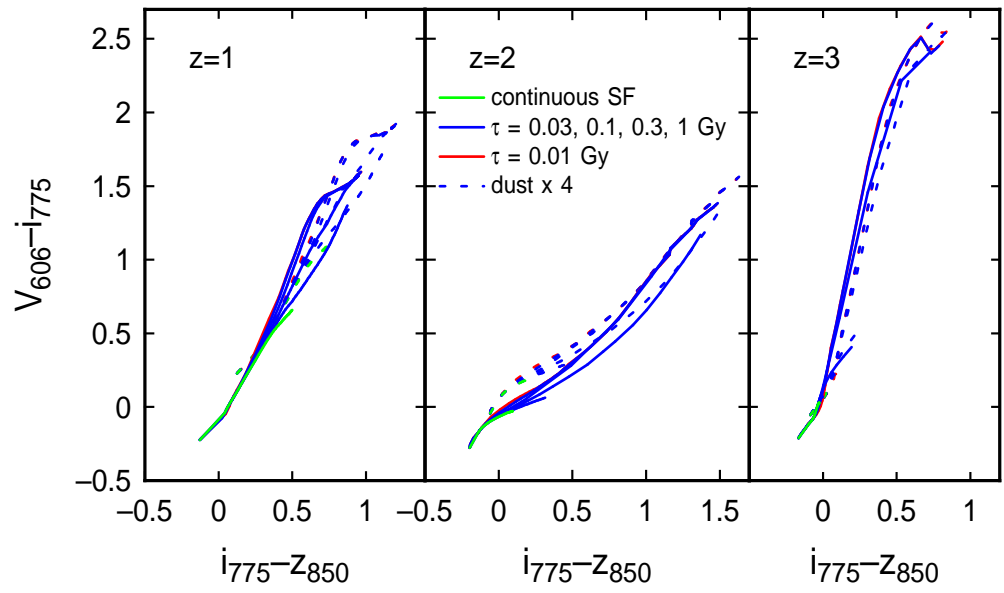}\caption{Model color-color
plots as in Figure 8. }\label{fig:viiz}\end{figure}

We solved for the three main variables (redshift, decay time, age)
in a certain way to minimize possible variations from intrinsic
scatter in the data.  We first solved for the minimum value of
$\delta=(bv-bv_m)^2+(vi-vi_m)^2+(iz-iz_m)^2$ along each curve that
traces out age for a given redshift and star formation history.
This minimum was found by a quadratic fit to the lowest three
$\delta$ values along that curve. Here, $bv$, $vi$, and $iz$
represent the observed colors and $bv_m$, $vi_m$ and $iz_m$
represent the model colors.  Next, the minimum values of $\delta$
were arranged in order of increasing redshift and, for each
redshift, increasing $\log(\tau)$ in the star formation history.
Along this list there were usually one or two broad minima in the
$\delta$ values, corresponding to close fits between the observed
and model colors.  Typically one close fit was at low $z$ and
another was at intermediate-to-high $z$. This duplicity
corresponds to the looping structure of the color-color curves in
Figure \ref{fig:bviz}, i.e., the template spectra have the same
slope at two distinct wavelengths (typically at the rest optical
and rest uv) for some star formation histories and extinctions. We
could rule out the small-$z$ solutions because these gave
effective rotation curve speeds (see below) that are unreasonably
low for disk-like galaxies (i.e., 20 km s$^{-1}$ or less).   The
broad minima along the ordered models were then isolated from each
other by considering only $\delta$ values less than certain
limits, 0.1, 0.05, and 0.02.  For each limit, the mean and the
standard deviation of the $z$ value was determined, weighted by
the inverse of $\delta$. When this mean $z$ was about the same for
each limit, and the standard deviation in $z$ became small,
$\pm0.2$ or less, for the smallest limit, then we felt confident
in our $z$ solutions.

The redshifts based on the whole galaxy colors were typically
several tenths smaller than the redshifts based on the average
clump colors. This difference is larger than the standard
deviation for each $z$ and also larger than the differences in
fitted $z's$ for the three limiting $\delta$ values.  Most likely,
the interclump emission, which contains about half the total
light, is contributing to the composite galaxy spectrum in a way
that makes the composite colors inconsistent with a single
exponential-decay model for star formation.  For example, whole
galaxies are redder than clumps in i$_{775}$-z$_{850}$, and the
models in Figure \ref{fig:bviz} shift i$_{775}$-z$_{850}$ toward
the red for lower $z$. We believe that the fits to the clump
colors give more accurate redshifts as long as we use simple star
formation histories.

For each fitted $z$, we also got the inverse-$\delta$ weighted
mean of $\log(\tau)$ in the star formation history and the
inverse-$\delta$ weighted mean of the age of the star formation
region, $t$.

\begin{figure}\epsscale{0.6}\plotone{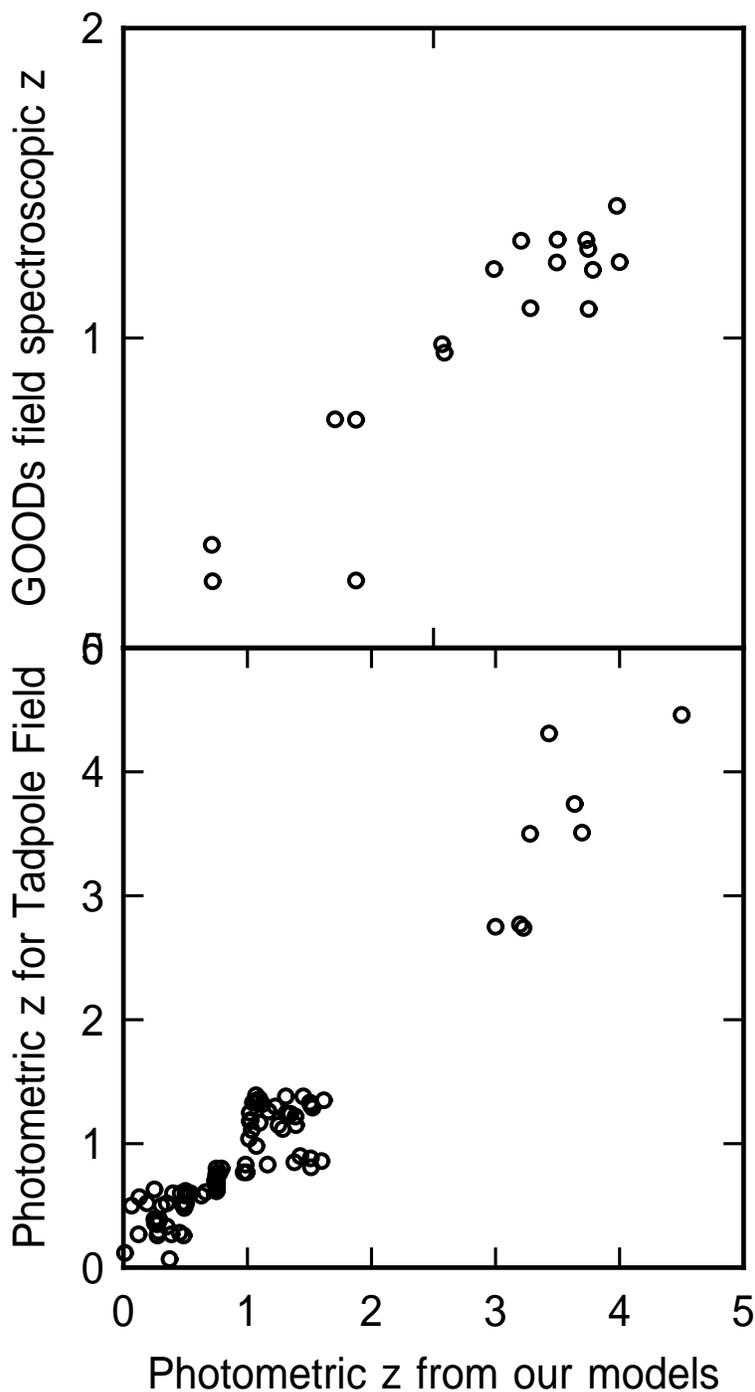}\caption{A comparison of redshifts
calculated by different methods. On the bottom are photometric
redshifts for the Tadpole Field galaxies calculated here
(abscissa) versus the Benitez et al. (2004) photometric redshifts
(ordinate). The filters appropriate for the Tadpole survey were
used; 29 galaxies with Benitez et al. redshifts in a narrow range
around 1.8 were omitted as they appear to be anomalous. On the top
are photometric redshifts for galaxies in the GOODS field
(abscissa) versus the Vanzella et al. (2004) spectroscopic
redshifts (ordinate). The filters appropriate to the GOODS survey
were used in this case; they are the same as the filters for the
UDF survey.} \label{redcheck}\end{figure}

To check our method, we compared the photometric redshifts derived
from our models with the 19 published redshifts available for the
GOODS field (Vanzella et al. 2004) using the photometry in that
study and the same filters as the UDF field. We also compared our
redshifts to those derived by Benitez et al. (2004) in the Tadpole
Galaxy field, using the different filters appropriate to the
Tadpole Galaxy survey. Figure \ref{redcheck} shows a plot of these
redshift distributions.  Our values matched the GOODS redshifts,
which range from 0.21 to 1.4, with an rms deviation of 0.2, not
counting an object at $z=5.83$, for which we derived $z=2.5$. This
single large discrepancy could result from strong emission lines
in the observation that are not present in the models, or from
inappropriate models at high $z$. Our values matched the 107
high-quality photometric redshifts in the Tadpole field (i.e.,
those for which ODDS$>0.99$) to within an rms of 0.3, not counting
29 more for which Benitez et al. derived a redshift of around
$1.85$. These excluded galaxies (not plotted) form a cluster in
the redshift distribution with a value of $z$ from Benitez et al.
$\sim1.872\pm0.015$; our redshifts for them range from 0.3 to 2.1,
with an average of $1.28\pm0.43$. Aside from these unexplained
discrepancies, our photometric redshifts are comparable to the
redshifts derived by others to within several tenths.  We also
determined redshifts for 30 B-band dropout galaxies in the UDF
field (Elmegreen et al. 2005) and obtained an average of
$3.49\pm0.37$, which is reasonable.

The interclump age could not be derived reliably from our models
because of the ambiguity between age and decay time $\tau$. The
interclump emission is probably a composite of many ages, however,
going back to the beginning of Pop II star formation, regardless
of the nature of the galaxies in which these stars formed. To find
the best interclump star formation histories, we used the galaxy
redshifts found from the clump colors and assumed an interclump
age corresponding to a beginning of interclump star formation at
$z=6$. We then determined the value of $\tau$ for each galaxy that
best fits the average three interclump colors.

\subsection{Redshift and age solutions}

Table \ref{tab:fits} shows the main results of the fitting
procedure with the Rowan-Robinson extinctions.  The average clump
age is derived to be 0.34 Gy from the clump colors and models. The
average interclump age is assumed to be the time between the
Universe age at the fitted clump redshift and the Universe age at
$z=6$, when the interclump star formation presumably began. The
average value in Table \ref{tab:fits} is 2.14 Gy. This larger age
for the interclump region is not unreasonable considering the
redder colors there. The clumps are characterized by a decaying
star formation rate with an e-folding time that has an average
$\tau\sim0.1$ Gy. Thus the clumps are three times older than their
e-folding times and should not be considered starbursts. The
fitted e-folding times for star formation in the interclump regions
average $\sim1$ Gy, which is only slightly less than the average
interclump age and consistent with a continued build-up of these
regions.

Extinctions E(B-V) larger than the Rowan-Robinson (2003) values by
factors of 2 and 4 for both clump and interclump emission give
about the same redshifts for the galaxies. The average values for
various quantities in these two cases are given in the bottom rows
of the tables.  The average redshift decreases from $z=2.3$ with
our fiducial extinction to 2.2 for both $2\times$ and $4\times$
extinctions. The higher extinctions decrease the average clump
ages from 0.34 Gy to 0.23 Gy and 0.09 Gy, respectively, with
$<\tau>=0.17$ and $0.16$ for the two cases, because the source has
to be intrinsically bluer when there is additional reddening. They
also make the interclump regions intrinsically bluer by increasing
the star formation decay times, $\tau$, from an average of 1.01 Gy
to averages of 2.09 Gy and 3.30 Gy, respectively.

The redshifts allow the determination of the luminosity distance
$d_L$ in equation \ref{eq:bc} (Carroll, Press, \& Turner 1992).
The resulting model apparent magnitudes scaled to the observed
apparent magnitudes give the clump masses.  This derivation of
mass uses the evolution models in GALAXEV again, as they give the
residual stellar mass per initial solar mass of a population of
stars as a function of time. To get the clump mass, we integrated
this residual stellar mass over time using the same time-dependent
star formation rate as for the integrated spectrum, and then
multiplied the resultant model mass by the factor used to scale
the model luminosity to the observed clump luminosity.

\subsection{Interclump mass column densities}

The interclump mass column densities were obtained from the
observed interclump surface brightnesses. To be careful about
surface brightness dimming, we first wrote the absolute magnitude
of a solar mass of stars from equation \ref{eq:bc} using $d_L=10$
pc. We then used the cosmological distance modulus (Carroll,
Press, \& Turner 1992; Paper III) to give the apparent magnitude
for this solar mass and scaled this result to the observed
apparent magnitude in a square arcsec. This gave the mass per
square arcsec. The square arcsec was then converted into a
square kiloparsec using the cosmological angle formula (Carroll,
Press, \& Turner 1992) to get the rest-frame mass column density.
This process corrects for the $\left(1+z\right)^4$ surface
brightness dimming. We got the same result using the $z$-dependent
luminosity distance $d_L$ in equation \ref{eq:bc}. The total
interclump mass was determined by averaging the interclump masses
per square arcsec for each galaxy and multiplying this average
by the galaxy area in square arcsecs. This assumes the
interclump medium is uniform. The interclump mass column density
is typically larger than the clump mass column density. Evidently,
the clump light is dominated by the brightest blue stars, while
the total mass in the region is dominated by the underlying old
stars. The sum of the clump mass and the interclump mass is taken
to be the luminous galaxy mass.

\subsection{Results}

Table \ref{tab:results} gives the average clump mass, $<M_{\rm
clump}>$, diameter $<D_{\rm clump}>$, and clump mass fraction,
$f_{\rm mass},$ in each clump-cluster galaxy, and it gives the
luminous galaxy mass, $M_{\rm gal}$, the galaxy diameter, $D_{\rm
gal}$ (equal to the major axis length out to the 2-$\sigma$
contour), and the effective rotation speed, $v_{\rm
rot}=\left[GM_{\rm gal}/\left(5D_{\rm gal}/2\right)\right]^{1/2}$
(the factor 5 in the denominator assumes a uniform spherical mass
distribution). For the physical sizes, $D$, the angular sizes were
used along with the cosmological angle-size relation for the
galaxy redshift.

The results in Table \ref{tab:results} may be summarized as
follows. The objects we have selected are large composite systems
with an average luminous mass of $6.5\times10^{10}$ M$_\odot$ and
an average diameter of 19.1 kpc out to the 2$\sigma$ contour. The
size and mass are not unprecedented for disk galaxies at redshifts
of 2 to 3 (Labb\'e et al. 2003; Shapley et al 2004). The average
rotation speed would be 150 km s$^{-1}$ if there were regular
circular motion, but there are no spectroscopic observations yet.
There is a range of a factor of 50 in mass and a factor of 8 in
rotation speed, so the peculiar morphology of this galaxy type
does not correspond to a particular mass. The clumps represent an
average total mass fraction of 19\%, again with a factor of 10
variation. This is smaller than their luminosity fraction
mentioned above (40\%) because the older interclump medium has a
higher mass-to-light ratio than the clumps.

The average interclump mass column density is $\sim4$ times larger
than the mass column density in solar neighborhood.  The primary
reason for the high value is the red color of the interclump
stars. Colors this red are matched to stellar populations with
high mass-to-light ratios. Increasing the assumed dust abundance
by factors of 2 and 4 for both the clumps and the interclump
regions hardly changed the redshifts and also changed the clump
masses only a little, from an average of $0.64\times 10^9$
M$_\odot$ to averages of $0.97\times 10^9$ M$_\odot$ and
$0.53\times 10^9$ M$_\odot$, respectively. These small changes
reflect the close balance between decreasing $M/L$ ratios for
intrinsically bluer stars and increasing $\exp\left(\tau_{\rm
dust}\right)$ corrections for more reddened stars.  The interclump
masses and total galaxy masses changed with extinction only a
little too: the average total galaxy masses changed from
$6.5\times10^{10}$ M$_\odot$ to $6.4\times10^{10}$ M$_\odot$
$11\times10^{10}$ M$_\odot$, respectively. Lower metallicity would
also make the intrinsic colors bluer, and this would make the
interclump ages older to match the observed colors, producing an
even higher $M/L$. Changes in the stellar initial mass function
could lower the inferred interclump mass, but only if the relative
abundance of stars younger than $\sim1$ Gy, corresponding to
A-type stars or earlier, is higher in proportion to low mass stars
than found locally; i.e., the IMF has to be relatively flat at
intermediate to high mass.

In another series of models, we assumed the Rowan-Robinson (2003)
extinction for the models of the clump emission, getting the same
clump redshifts and ages as before, but then we assumed higher
extinctions in the interclump regions by factors of 2 and 4 before
deriving their star formation histories and masses.  These higher
interclump extinctions gave intrinsically bluer colors and a
greater extinction correction to the magnitudes. The resultant
interclump masses averaged for all 10 galaxies increased from
$5.9\times10^{10}$ M$_\odot$ to $6.5\times10^{10}$ M$_\odot$
$12\times10^{10}$ M$_\odot$ respectively, as the corresponding
lower mass-to-luminosity ratios were more than offset by a higher
mass needed to overcome the greater extinction.

Other clump properties are in Table \ref{tab:other}.  The clump
equivalent atomic density, $n_{\rm clump}$, is taken to be the
ratio of the clump stellar mass to the volume given by the clump
diameter, $D_{\rm clump}$, assuming a mean molecular weight
$2.2\times10^{-24}$g.  This equivalent density is useful for
comparisons with molecular cloud and other gas densities (see
below).  The dynamical time is $t_{\rm
dyn}=\left(G\rho\right)^{-1/2}$ for mass density $\rho$. The
average and current star formation rates (SFR) are given by
\begin{equation}
{\rm SFR(ave)} = M_{\rm clump}/t\;\;;\;\;{\rm SFR}={{M_{\rm
clump}e^{-t/\tau}}\over
{\tau\left(1-e^{-t/\tau}\right)}}\end{equation} for clump age $t$
and decay time $\tau$ in Table \ref{tab:fits}.  The internal clump
velocity dispersion assumes virial equilibrium with a uniform
density: $\Delta v_{\rm clump}= \left[GM_{\rm
clump}/\left(2.5D_{\rm clump}\right)\right]^{1/2}$. The last
column is the ratio of the clump internal dynamical time to the
galaxy dynamical time. All of these properties are discussed
below.

\section{Discussion}

\subsection{General considerations}

Clump-cluster galaxies are a new type of object seen primarily in
deep fields such as those surveyed by HST. They are characterized
by: (1) a disk-like shape (Paper II); (2) a significant
contribution to the total luminosity ($\sim40$\% -- Table
\ref{tab:prop}) from a small number of relatively blue clumps; (3)
a flat light profile along the major axis (Papers I, II); (4) no
central red clump or bulge (Papers I, II); and (5) a faint,
relatively red interclump emission (this paper). Some
clump-clusters also have bar-like features (Paper III). Edge-on
clump-clusters may be chain galaxies (Paper II).  The lack of
clear rotation in chain galaxies (Bunker et al. 2000; Erb et al.
2004) is a mystery if these are normal galaxy disks.

In the few cases studied here, the clumps are typically $\sim0.3$
Gy old in their rest frame and not currently forming stars as fast
as they once did.   Emission from the interclump medium is
dominated by stars that are much older, perhaps up to 2 Gy older
in some cases. Older and fainter populations inside the clumps
could be present but undetected.

With $\sim19$\% of the luminous mass in a dozen or so giant
clumps, these galaxies are unlike modern galaxies where the
largest star forming regions are typically $<10^{-4}$ of a
galaxy's mass. This difference implies several things. First, the
extreme clumpiness suggests that the rotation curve will not be
simple (Immeli et al. 2004a,b). The clumps should attract each
other and scatter, with their coldest and most dissipative parts
migrating to the center to make a bulge and their hottest, least
dissipative parts filling the interclump region or thick disk
(Noguchi 1999; Somerville, Primack \& Faber 2001; Immeli et al.
2004a,b; Hartwick 2004; Brook et al. 2004). The lack of clear
rotation in the few chain galaxies that have been observed may be
the result of this severe clump interaction.

Second, the clump ages are longer than their dynamical times, so
the clumps are self-bound like giant clusters. The clump density
is comparable to the mean galaxy density, however, so tidal forces
on the clumps are significant (this follows from the comparable
clump and galaxy dynamical times in Table \ref{tab:other}). This
means that galactic shear and tidal stretching are moderately
important for the lowest density clumps. For example, UDF
6462+6886 and UDF 82+99+103 have clumps with tails and arcs
suggesting the influence of tidal forces.

The fact that the equivalent clump atomic densities are much less
than typical molecular cloud densities implies that the clump
stars probably formed in numerous dense cores inside kpc-size,
gravitationally-bound, low-density cloud complexes. This
hierarchical morphology for star formation is analogous to the
case of star complexes found in local spiral galaxies (Efremov
1995), except that for the clump clusters, the clumps are more
strongly bound; local star complexes readily disperse or shear
into flocculent spiral arms.

The internal velocity dispersions in the clumps are inferred to be
moderately large, 24 km s$^{-1}$ on average, based on the virial
theorem.  This velocity is $\sim16$\% of the average galactic
orbital speed -- a high fraction for gas in local spirals but
comparable to the fraction in dwarf galaxies. If the clouds that
made the young stellar clumps formed by gravitational
instabilities (Noguchi 1999; Immeli et al. 2004a,b) and if they
still have the ambient Jeans mass, then the velocity dispersion in
the pre-cloud gas had to be comparable to the virial velocity in
the clumps (this is based on the usual isothermal assumption).
Such a high dispersion implies that clump-cluster galaxies are far
more turbulent than local galaxies.

One source of turbulence could be major encounters with other
galaxies. Tidal forces increase the velocity dispersion
considerably in local galaxies undergoing such encounters (Kaufman
et al. 1997).  The galaxies in our sample have no major companions
now, but the giant clumps in them could have been triggered
$\sim0.3$ Gy ago at previous encounters, considering the average
clump age. Another possible source of turbulence is accretion from
the intergalactic medium. The disk impact speed can be $\sim0.4$
times the circular speed if the gas falls in along cosmological
filaments (Keres et al. 2004). This would be $\sim60$ km s$^{-1}$
for our galaxies. Disk turbulence should be slower than this
because of energy dissipation, but still enough to explain the
high $\Delta v_{clump}$.  Supernovae could also be an important
source of turbulence as they are in local galaxies, but in
clump-cluster galaxies they are likely to be highly concentrated
in the clumps, and then they may just cause blow-out rather than
widespread turbulence.

In the accretion model, clump-cluster galaxies would have added
intergalactic gas at a rate comparable to the average star
formation rate, $\sim 20$ M$_\odot$ yr$^{-1}$.  The disks would
presumably have become unstable when the mass column density
exceeded a critical value. Considering the clump ages, the
resulting star formation has a timescale of $\sim0.3$ Gy, which is
several galactic orbit times.

\subsection{Accretion models}

A different possibility is that the clumps were formerly separate
galaxies that accreted during hierarchical buildup (e.g., Abadi et
al. 2003; Brook et al. 2004). Some of our clump clusters are so
irregular in morphology (e.g., UDF 6462+6886) that they appear
more like random conglomerations than circular disks. Also, the
lack of clear rotation in the few cases observed suggest highly
disturbed dynamics. This model raises three questions: (1) Can
small galaxies like our clumps survive their capture by large
galaxies, considering the interclump population that was present
before? (2) Are there examples of bare clumps in the UDF that
resemble the clumps in our sample? (3) Would the distribution of
axial ratios for coalesced systems be the same as that observed
for clump clusters and chain galaxies? We attempt to answer these
questions here.

\subsubsection{Clump Survival}

The short-term survival of a small galaxy captured by a large
galaxy depends mostly on the ratio of densities.  This ratio is
important because it is also the ratio of the self-binding force
per unit length inside the small galaxy to the tidal disruptive
force per unit length exerted by the large galaxy. If the small
galaxy is much denser than the large galaxy, then the small galaxy
will survive until it migrates, perhaps because of dynamical
friction, to a sufficiently dense region of the large galaxy, at
which time it disperses.  The migration time depends on the orbit,
and the outer parts of the clump can be shredded continuously as
it migrates. In the clump-cluster galaxies studied here, the
typical column density of a clump is comparable to the average
column density of the interclump population, both being several
hundred M$_\odot$ pc$^{-2}$.  The clump density should be
considerably higher than the interclump density, however, because
the interclump medium is likely to be much thicker than any single
clump as a result of stirring from clump motions, dissolution of
previous clumps, and stellar migration. The ratios of axes for
these galaxies also suggest fairly thick disks (Paper II, and see
below). In addition, the clumps probably have dense cores like
other bound clusters. The clumps are therefore likely to survive
their first few impacts with the disk, but probably not much
longer. The observation that the clumps are much older than both
their internal dynamical times and the galaxy orbit times (by a
factor of 10) implies that most of them live outside the densest
regions of the disk.  The exceptions are the clumps with
morphologies suggesting tidal disruption, such as the head-tail
shapes of the clumps in UDF 82+99+103, UDF 6462+6886, and UDF
3034+3129.

\subsubsection{Bare clumps in the UDF}

\begin{figure}\epsscale{0.8}\plotone{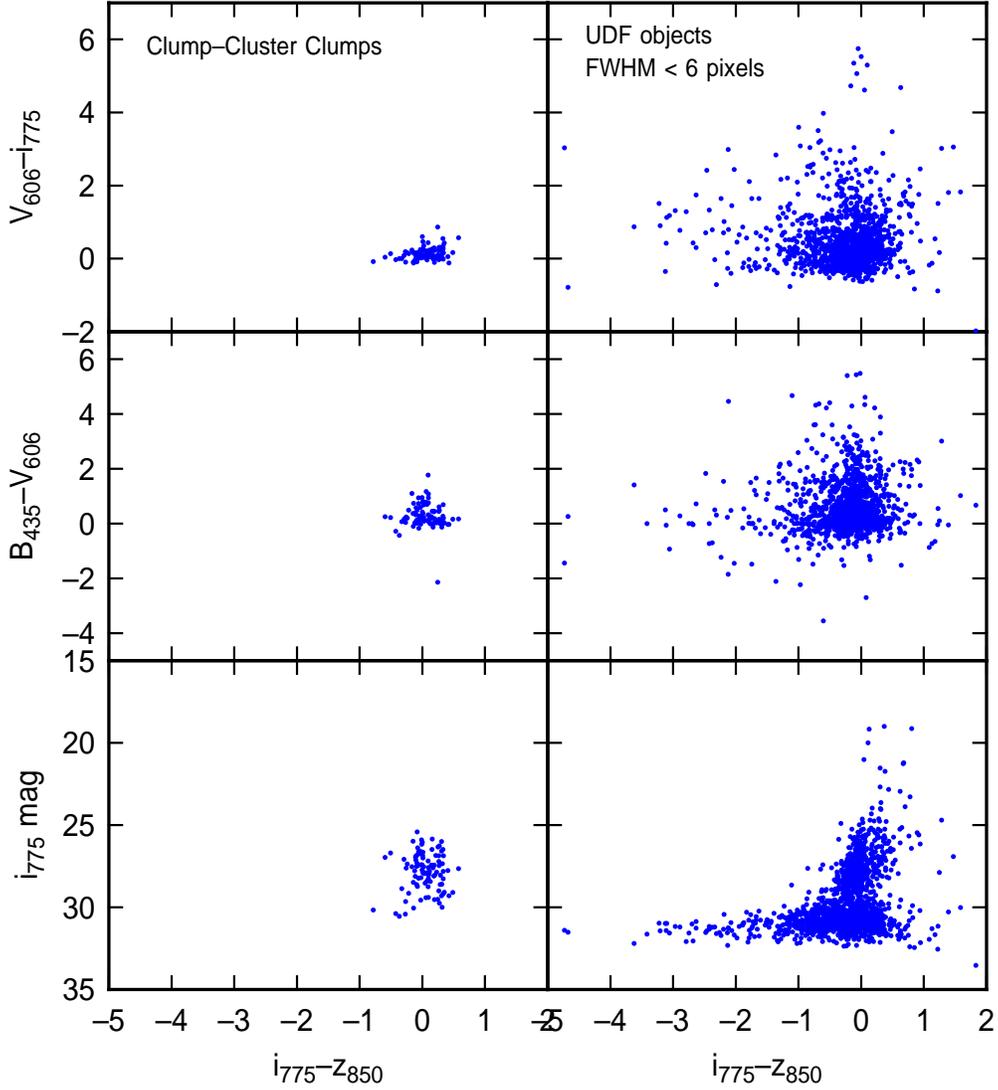}\caption{(left)
Color-color and color-magnitude distributions for all the measured
clumps in our 10 clump-cluster galaxies, and (right) analogous
distributions for all the small clumps in the UDF field, with
diameters less than 6 pixels. The similar distribution centroids
suggest that the clumps in our clump-cluster galaxies could have been
captured from the field. The faintest objects in the field also have
faint counterparts in clump-cluster galaxies, but they were not
included in our clump measurements.}
\label{fig:udfsmall}\end{figure}

What is the evidence for clumps like these outside of
clump-cluster galaxies, where they presumably existed before the
accretion? The UDF contains many isolated objects that resemble
our clumps in both luminosity and color. The right-hand side of
Figure \ref{fig:udfsmall} shows color-magnitude and color-color
plots of all the UDF objects smaller than 6 pixels in FWHM, as
given in the tabulation on the UDF web
site\footnote{http://archive.stsci.edu/pub/hlsp/udf/acs-wfc/h\_udf\_wfc\_VI\_I\_cat.txt
}. This size limit was chosen because it roughly corresponds to
the size of a clump in a clump-cluster galaxy. The left side of
the figure shows the magnitudes and colors for the measured clumps
in the clump-cluster galaxies of this paper. The distributions are
essentially the same in the regions of overlap. The UDF clumps can
be much bluer than our clumps, suggesting either that star
formation begins to slow down once a clump is ingested into a
larger galaxy, or the clump-cluster sample in our survey has too
few clumps to include the rare active ones seen in the general UDF
field. There are also much fainter clumps in the UDF field than
those measured in our survey, but the clump-cluster galaxies have
much fainter clumps too which we did not study. We note that these
small objects in the UDF field are not the well-studied Lyman
Break galaxies, which tend to be more massive, blue, and luminous
than our clumps even at the same $z\sim2.5-3$ (e.g., Papovich,
Dickinson, \& Ferguson 2001). The small field objects are more the
size of the $z\sim6$ galaxies studied by Bunker et al. (2004).
Most likely, most are low-mass and low-luminosity galaxies in
about the same range of redshifts as the clump-cluster galaxies
studied here.

\subsubsection{Distribution of axial ratios}

The distribution of axial ratios for a random assembly of field
clumps might be expected to be significantly different than the
distribution for a thick disk.  This distribution was determined
for a combined sample of chain and clump-cluster galaxies in Paper
II and shown to be similar to the distribution for nearby spiral
galaxies (see also O'Neil et al. 2000; Reshetnikov, Dettmar, \&
Combes 2003).  In general, a thin disk should have a flat
distribution for the ratio of minor to major axes, $W/D$, with a
slight rise and then a drop at low ratios where the projected
width $W$ becomes comparable to the intrinsic disk thickness.  A
galaxy whose appearance is dominated by a few clumps will not be
very different, in fact.  The first three clumps make a plane
regardless of their positions, so the question is how much do the
few additional clumps in a clump-cluster or chain galaxy distort
this projected plane?

\begin{figure}\epsscale{0.6}\plotone{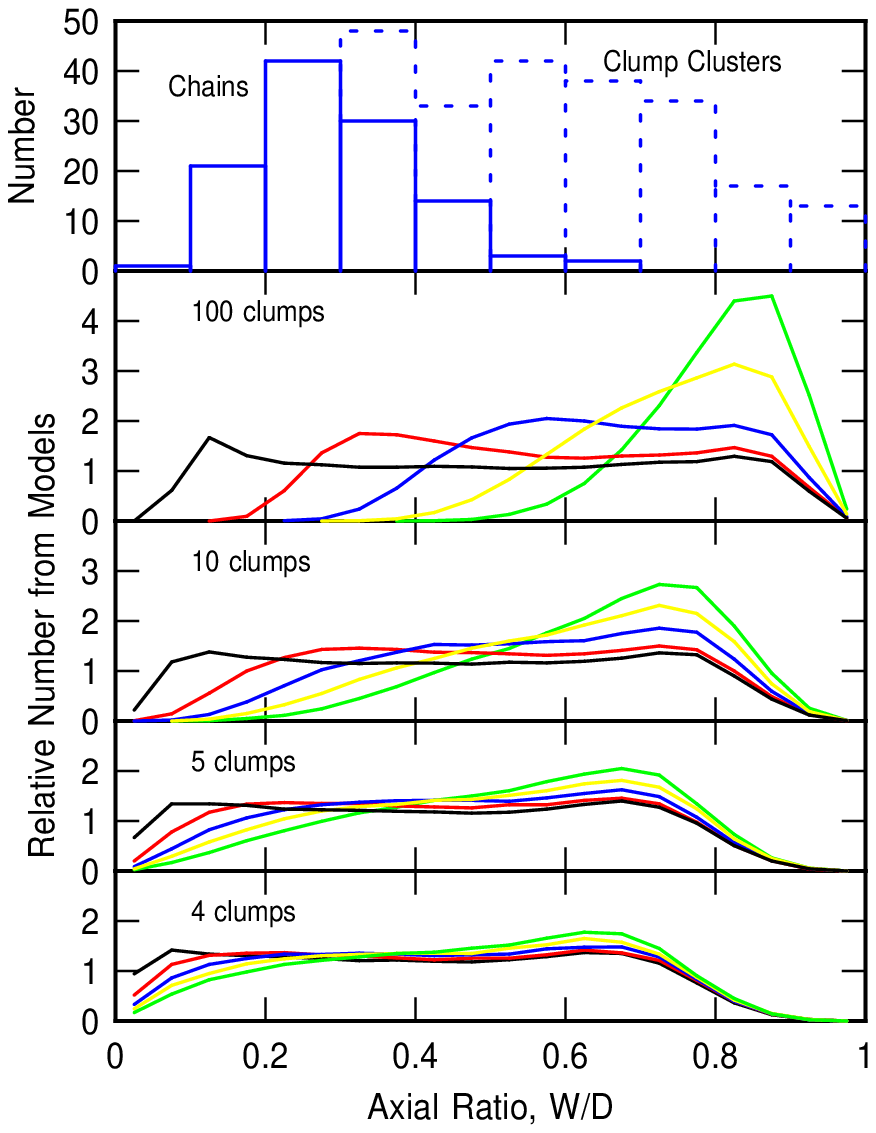}\caption{(top) The distribution of
the ratio of minor to major axes for all the chain galaxies (solid
line) and clump-cluster galaxies (incrementally represented by the
dashed line) in our larger survey of the UDF field. The bottom 4
panels show models of axial ratios for pure-clump galaxies with
randomly positioned clumps. In each case, there are 5 curves that
correspond to intrinsic flattenings of $Z=0.1$, 0.3, 0.5, 0.7 and
1 (which is an unflattened spherical distribution). These curves
move progressively to the right in the panels. The number of
clumps in the models is different for each panel, as indicated.
The figure shows how UDF galaxies dominated by 5 to 10 clumps,
such as chain and clump-cluster galaxies, are most likely
flattened with an intrinsic axial ratio of 0.1 to 0.3, as shown by
the black and red curves (in the electronic edition) in the second
and third panels up from the bottom. The decrease in the
distribution at high axial ratio is from the intrinsically
irregular structure of the galaxies (smooth disks would not have
such a decrease), while the decrease at low axial ratio is from
the minimum intrinsic flattening. Evidently, the clump-cluster galaxies
are in the process of forming thick disks.
}\label{fig:clumps}\end{figure}

Figure \ref{fig:clumps} shows distribution functions of axial
ratios for model galaxies with various numbers of clumps randomly
positioned in 3 dimensions. Each line is a histogram from $10^5$
models. The first three clumps in each model are taken to define a
plane and the others are given random positions around the plane
in polar coordinates. The radial distance of each clump is
distributed as a Gaussian. To simulate variations from different
degrees of flattening in a disk, the distance perpendicular to the
plane is multiplied by factors of $Z=0.1,$ 0.3, 0.5, 0.7, and 1
for the five different lines.  Once a clump distribution is
chosen, the polar axis of the model is tilted by a random angle
whose cosine has a uniform distribution function, simulating
random inclinations of the model in the sky. The axial ratio used
for the histogram is then taken to be the minimum ratio of
projected width to length, considering all angles in the sky
plane. The resulting histograms peak for axial ratios near unity
when $Z=1$ and there are a lot of clumps, because then the model
galaxy is a uniformly filled sphere. When the number of clumps is
small, the distribution decreases at high axial ratios because the
clumps are irregularly positioned even in the spherical case
($Z=1$) and some directions happen to have have short sizes. The
distributions decrease at small axial ratio because there are
usually some clumps at non-zero perpendicular distance, making
$W/D$ non-zero. The axial ratio where this decrease begins is
smaller for more flattened systems because the minimum value of
$W/D$ is $\sim Z$.

The observed distribution of axial ratios for all of the chains
and clump-cluster galaxies in our survey of the UDF field
(Elmegreen et al. 2005) is also shown in Figure \ref{fig:clumps}.
The solid line is for chains and the dashed line is for
clump-clusters, added on top of the solid line. The observed
distribution resembles the models only when there is some
intrinsic flattening of the galaxies. However, the degree of
flattening does not have to be large.  The best models are for a
number of clumps appropriate to our galaxies, $\sim5-10$, and for
a flattening $Z\sim0.3$, which is the second curve from the left
(red in the electronic version).  The corresponding intrinsic
axial ratio of the model galaxies is the same, $\sim0.3$, which
implies the disks are very thick.  The same conclusion was reached
in Paper II.

\subsection{Assessment of accretion model}

These considerations suggest that the clumps in some of our
clump-cluster galaxies could have been accreted as whole objects
from the field. They would probably survive for the required time,
there are isolated field galaxies like them, and the
clump-clusters are relatively thick disks. Some energy dissipation
perpendicular to the disk had to occur to flatten the clump
distribution to the observed axial ratio of 0.1 to 0.3, but
dynamical friction from the observed interclump disk could have
done this during the capture process.

This capture scenario is consistent with the model by Brook et al.
(2004), who suggest that thick disks in spiral galaxies form
quickly, in less than 1 Gy, as a result of mergers with a small
number of $\sim2\times10^9$ $M_\odot$ clumps. Hartwick (2004) also
notes the quick formation time of thick disks, and suggests that
the pre-galactic clumps which make them could remain today, in
part, as disk globular clusters.

A peculiarity of our sample is the high interclump surface
density. If this result is correct, then it implies that the
galaxies cannot simply age to form present-day spiral disks.
Today's disks either have high mass column densities along with
massive bulges, or low mass column densities without prominent
bulges. They do not have high mass column densities and no bulges,
like our clump-clusters. One possibility is that a large fraction
of the mass in clump-cluster disks will migrate inward to form a
bulge. This migration would be accompanied by outward angular
momentum transfer, and might result from spiral waves or repeated
interactions, with bar formation a possible intermediate step.
Another possibility is that clump-cluster galaxies merge to form
elliptical galaxies. The likelihood of this is high considering
their relatively large sizes and high redshifts where interactions
are still frequent.

B.G.E. is supported by the National Science Foundation
through grant AST-0205097.  Support for D.M.E. was provided by
the Salmon Fund of Vassar College.

\clearpage

\begin{deluxetable}{llccccccc}
\tabletypesize{\scriptsize} \tablecaption{Properties of Galaxies
\label{tab:prop}}
\tablewidth{0pt}
\tablehead{
\colhead{UDF ID Numbers} &
\colhead{Position\tablenotemark{a}} &
\colhead{i$_{775}$} &
\colhead{B$_{435}$-V$_{606}$} &
\colhead{V$_{606}$-i$_{775}$ } &
\colhead{i$_{775}$-z$_{850}$ } &
\colhead{$\mu_{775}$} &
\colhead{No.} &
\colhead{$f_{\rm lum}$}\\
\colhead{} &
\colhead{$3^{\rm h}$ $32^{\rm m}, $ $-27^\circ$} &
\colhead{mag} & \colhead{} &
\colhead{} &
\colhead{} &
\colhead{mag arcsec$^{-2}$} &
\colhead{} &
\colhead{} }
\startdata
82+99+103&  $40.7472^{{\rm s}},$  $49^\prime$ $26.047^{\prime\prime}$ &   23.6  &  0.43  &  0.25  &  0.30   & 25.1 & 8  & 0.48\\
3034+3129&   $37.8404^{{\rm s}},$  $47^\prime$  $56.390^{\prime\prime}$ &   24.0  &  0.82  &  0.18  &  0.08   & 24.6 & 5  & 0.44\\
2012&   $35.9178^{{\rm s}},$  $48^\prime$  $17.216^{\prime\prime}$ &   24.3  &  0.80  &  0.11  &  0.12   & 25.0 & 10 & 0.38\\
3483&   $36.3842^{{\rm s}},$  $47^\prime$  $47.125^{\prime\prime}$ &   23.4  &  0.20  &  0.14  &  0.13   & 24.5 & 12 & 0.30\\
3465+3597&   $35.6573^{{\rm s}},$  $47^\prime$  $48.567^{\prime\prime}$ &   23.5  &  0.11  &  0.04  &  0.02   & 24.3 & 11 & 0.34\\
CC-1\tablenotemark{b}&   $49.3154^{{\rm s}},$  $47^\prime$  $26.377^{\prime\prime}$ &   22.0  &  0.44  &  0.16  &  $-0.07$  & 23.9 & 11 & 0.13\\
4245+4450+&   $28.9470^{{\rm s}},$  $47^\prime$  $37.730^{\prime\prime}$ &   22.6  &  0.08  &  2.75  &  $-2.18$  & 23.7 & 14 & 0.06\\
\hspace{.2cm}4466+4501+4595 &&&&&&&&\\
1666&   $41.3128^{{\rm s}},$  $48^\prime$  $21.154^{\prime\prime}$ &   23.9  &  0.09  &  0.10  &  0.45   & 25.2 & 13 & 0.39\\
7081+7136+7170&   $38.9909^{{\rm s}},$  $46^\prime$  $56.265^{\prime\prime}$ &   24.5  &  0.06  &  0.10  &  0.34   & 25.9 & 8  & 0.38\\
6462+6886&   $43.5596^{{\rm s}},$  $46^\prime$  $59.376^{\prime\prime}$ &   23.5  &  0.18  &  0.32  &  0.27   & 25.5 & 8  & 0.36\\
\enddata
\tablenotetext{a}{only the seconds of Right Ascension and the arc minutes and seconds of Declination are tabulated.}
\tablenotetext{b}{no UDF number given for this because it is near the edge of the image,
so we use CC for "Clump Cluster".}
\end{deluxetable}

\begin{deluxetable}{lccccc}
\tabletypesize{\scriptsize} \tablecaption{Redshift and Star Formation Properties
\label{tab:fits}} \tablewidth{0pt}
\tablehead{
\colhead{Galaxy} &
\colhead{redshift} &
\colhead{Clump} &
\colhead{Clump} &
\colhead{Interclump}&
\colhead{Interclump} \\
\colhead{Number} &
\colhead{z\tablenotemark{a}} &
\colhead{age, t (Gy)} &
\colhead{decay $\tau$ (Gy)\tablenotemark{b}} &
\colhead{age (Gy)\tablenotemark{c}}&
\colhead{decay $\tau$ (Gy)}
}
\startdata

82+&   2.5&    0.36&   0.11&   1.71&   0.33\\
3034+&   3.0&    0.21&   0.10&   1.24&   0.58\\
2012&   3.0&    0.12&   0.03&   1.26&   0.65\\
3483&   2.2&    0.26&   0.05&   2.02&  1.63\\
3465+&   2.4&    0.22&   0.06&   1.80&  1.06\\
CC-1&   2.8&    0.35&   0.14&   1.40&  1.38\\
4245+&   1.7&    0.32&   0.05&   2.92&  1.16\\
1666&   1.6&    0.81&   0.28&   3.15&   0.91\\
7081+&   1.6&    0.43&   0.10&   3.13&   0.92\\
6462+&   1.8&    0.31&   0.06&   2.82&  1.49\\
averages =&   2.3&    0.34&   0.10&   2.14&  1.01\\
ave, $A_V\times2$ =&   2.2&    0.23&   0.17&   2.40&  2.09\\
ave, $A_V\times4$ =&   2.2&    0.09&   0.16&   2.32&  3.30\\

\enddata
\tablenotetext{a}{Photometric redshifts for the average clump
colors} \tablenotetext{b}{Average star formation history
exponential decay rate for clumps}\tablenotetext{c}{Assumes
interclump star formation began at $z=6$.}
\end{deluxetable}

\begin{deluxetable}{lcccccc}
\tabletypesize{\scriptsize} \tablecaption{Derived Properties
\label{tab:results}} \tablewidth{0pt}
\tablehead{
\colhead{Galaxy} &
\colhead{$<M_{\rm clump}>$} &
\colhead{$<D_{\rm clump}>$}&
\colhead{$f_{\rm mass}$}&
\colhead{$M_{\rm gal}$} &
\colhead{$D_{\rm gal}$} &
\colhead{$v_{\rm rot}$}\\
\colhead{Number} &
\colhead{$10^9$ M$_\odot$} &
\colhead{kpc} &
\colhead{} &
\colhead{$10^{10}$ M$_\odot$} &
\colhead{kpc} &
\colhead{km s$^{-1}$} }
\startdata
82+&      1.36&       2.5&       0.04&     25&      18&   350\\
3034+&       0.70&       2.1&       0.16&      2.2&      16&   110\\
2012&       0.17&       1.5&       0.07&      2.5&      13&   130\\
3483&       0.75&       1.4&       0.32&      2.8&      14&   130\\
3465+&       0.50&       1.5&       0.28&      2.0&      17&   100\\
CC-1&       0.96&       1.3&       0.07&     15&      25&   230\\
4245+&       0.21&       2.1&       0.59&       0.5&      22&    44\\
1666&       0.52&       1.6&       0.08&      8.7&      19&   200\\
7081+&       0.35&       1.9&       0.11&      2.6&      20&   110\\
6462+&       0.90&       2.4&       0.22&      3.3&      29&    100\\
average =&       0.64&       1.8&       0.19&       6.5&      19&   150\\
ave, $A_V\times2$ =&       0.97&       1.8&       0.10&       6.4&      19&   140\\
ave, $A_V\times4$ =&       0.53&       1.9&       0.07&      11&      19&   180\\

\enddata

\end{deluxetable}

\begin{deluxetable}{lcccccc}
\tabletypesize{\scriptsize}\tablecaption{Other Clump Properties
\label{tab:other}}
\tablewidth{0pt}
\tablehead{
\colhead{Galaxy} &
\colhead{$n_{\rm clump}$} &
\colhead{$t_{\rm dyn}$ } &
\colhead{SFR(ave) } &
\colhead{SFR(now) } &
\colhead{$\Delta v_{\rm clump}$  } &
\colhead{$t_{\rm dyn}v_{rot}/D_{\rm gal}$} \\
\colhead{Number}&
\colhead{cm$^{-3}$}&
\colhead{Gy}&
\colhead{M$_\odot$ yr$^{-1}$}&
\colhead{M$_\odot$ yr$^{-1}$}&
\colhead{km s$^{-1}$}&
\colhead{}
}
\startdata
82+&       5.1&      0.037&      31&      3.9&       31&       0.73\\
3034+&       3.3&      0.046&      17&      4.0&       24&       0.32\\
2012&       2.6&      0.052&      14&      1.1&       14&       0.53\\
3483&      16&      0.021&      34&       0.6&       30&       0.20\\
3465+&       7.4&      0.030&      25&      2.6&       24&       0.19\\
CC-1&      24&      0.017&      31&      7.0&       35&       0.15\\
4245+&       1.1&      0.081&       9.3&       0.1&       13&       0.17\\
1666&       5.0&      0.037&       8.4&      1.4&       24&       0.39\\
7081+&       2.4&      0.053&       6.5&      0.4&       18&       0.28\\
6462+&       3.7&      0.043&      23&       0.5&       25&       0.15\\
average =&       7.0&      0.042&      20&      2.2&       24&       0.31\\
ave, $A_V\times2$ =&  6.7&      0.068&      25&      4.7&       22&      0.38\\
ave, $A_V\times4$ =&       2.6&      0.090&      42&     30&       15&       0.79\\

\enddata

\end{deluxetable}

\newpage

%fig1

%fig2

%fig3

%fig4

%fig5

%fig6

%fig7

%fig8

%fig9

%fig10

%fig11

%fig12


\begin{thebibliography}{}

\bibitem[]{791} Abadi, M.G., Navarro, J.F., Steinmetz, M., Eke, V.R.
2003, ApJ, 597, 21

\bibitem[]{794}Abraham, R., van den Bergh, S., Glazebrook, K.,
Ellis, R., Santiago, B., Surma, P., \& Griffiths, R. 1996, ApJS,
107, 1

\bibitem[]{798} Adelberger, K.L., \& Steidel, C.C. 2000, ApJ, 544,
218

\bibitem[]{801} Beckwith, et al. 2004, preprint

\bibitem[]{} Benitez, N. et al. 2004, ApJS, 150, 1

\bibitem[]{803} Brook, C.B., Kawata, D., Gibson, B.K., \& Freeman,
K.C. 2004, ApJ, 612, 894

\bibitem[]{806} Bruzual, G. \& Charlot, S. 2003, MNRAS, 344, 1000

\bibitem[]{808} Bunker, A., Spinrad, H., Stern, D., Thompson, R.,
Moustakas, L., Davis, M., \& Dey, A. 2000, in Galaxies in the
Young Universe II, ed. H. Hippelein \& K. Meisenheimer (Berlin:
Springer)

\bibitem[]{813} Calzetti, D., Armus, L., Bohlin, R.C., Kinney, A.L.,
Koornneef, J., \& Storchi-Bergmann, T.  2000, ApJ, 533, 682

\bibitem[]{594} Carroll, S.M., Press, W.H., \& Turner, E.L. 1992,
ARAA, 30, 499

\bibitem[]{819} Chabrier, G. 2003, PASP, 115, 763

\bibitem[]{821} Conselice, C. 2004, astro-ph/0405102

\bibitem[]{823} Conselice, C., Grogin, N.A., Jogee, S., Lucas, R.A.,
Dahlen, T., de Mello, D., Gardner, J.P., Mobasher, B.,
Ravindranath, S. 2004, ApJ, 600, L139

\bibitem[]{827} Conselice, C.J., Blackburne, J.A., \& Papovich, C.
2004, astroph/0405001

\bibitem[]{830} Cowie, L., Hu, E., \& Songaila, A. 1995, AJ, 110,
1576

\bibitem[]{} Dalcanton, J.J. \& Bernstein, R.A. 2002, AJ, 124,
1328

\bibitem[]{833} Dalcanton, J.J., \& Schectman, S.A. 1996, ApJ, 465,
L9

\bibitem[]{836} Dickinson, M., Papovich, C., Ferguson, H.C., \&
Budav\'ari, T. 2003, ApJ, 587, 25

\bibitem[]{} Efremov, Y.N. 1995, AJ, 110, 2757

\bibitem[]{839}Elmegreen, D.M., Elmegreen, B.G., \& Sheets, C. 2004,
ApJ, 603, 74 (Paper I)

\bibitem[]{842}Elmegreen, D.M., Elmegreen, B.G., \& Hirst, A.C.
2004a, ApJ, 604, L21 (Paper II)

\bibitem[]{845} Elmegreen, B.G., Elmegreen, D.M., \& Hirst, A.C.
2004b, ApJ, 612, 191 (Paper III)

\bibitem[]{848} Elmegreen, D.M., et al. 2005, in preparation

\bibitem[]{850} Erb, D.K., Steidel, C.C., Shapley, A.E., Pettini,
M., \& Adelberger, K.L. 2004, 612, 122

\bibitem[]{853} Ferguson, H. C., Dickinson, M., \& Williams, R.
2000, ARA\&A, 38, 667

\bibitem[]{856} Hartwick, F.D.A. 2004, ApJ, 603, 108

\bibitem[]{858} Idzi, R. et al. 2004, ApJ, 600, L115

\bibitem[]{860} Immeli, A., Samland, M., Gerhard, O., \& Westera, P.
2004a, A\&A, 413, 547

\bibitem[]{863} Immeli, A., Samland, M., Westera, P., \& Gerhard, O.
2004b, ApJ, 611, 20

\bibitem[]{866} Iono, D., Yun, M.S., \& Mihos, J.C. 2004, ApJ, 616,
199

\bibitem[]{869} Kaufman, M., Brinks, E., Elmegreen, D. M.,
Thomasson, M., Elmegreen, B. G., \& Struck, C., 1997, AJ, 114,
2323

\bibitem[]{873} Keres, D., Katz, N., Weinberg, D.H., \& Dav\'e, R.
2004, MNRAS, in press, astroph/0407095

\bibitem[]{876} Labb\'e, I., et al. 2003, ApJ, 591, L95

\bibitem[]{878} Leitherer, C., Li, I.-H., Calzetti, D., Heckman,
T.M. 2002, ApJS, 140, 303

\bibitem[]{881} Madau, P. 1995, ApJ, 441, 18

\bibitem[]{883} Moustakas, L. et al. 2004, ApJ, 600, L131

\bibitem[]{885} Murali, C., Katz, N., Hernquist, L., Weinberg, D.H.,
\& Dav\'e, R. 2002, ApJ, 571, 1

\bibitem[]{888} Noguchi, M. 1996, ApJ, 514, 77

\bibitem[]{890} O'Neil, K., Bothun, G.D., \& Impey, C.D. 2000, ApJS,
128, 99

\bibitem[]{893} Papovich, C., Dickinson, M., \& Ferguson, H.C. 2001,
ApJ, 559, 620

\bibitem[]{896}Reshetnikov, V., Dettmar, R.-J., \& Combes, F. 2003,
A\&A, 399, 879

\bibitem[]{899} Roche, N., Ratnatunga, K., Griffiths, R. E., \& Im,
M. 1997, MNRAS, 288, 200

\bibitem[]{902} Rowan-Robinson, M. 2003, MNRAS, 345, 819

\bibitem[]{904} Rudnick, G. et al. 2003, ApJ, 599, 847

\bibitem[]{906} Shapley, A.E., Steidel, C.C., Adelberger, K. L.,
Dickinson, M., Giavalisco, M., Pettini, M. 2001, ApJ, 137, 139

\bibitem[]{909} Shapley, A.E., Erb, D.K., Pettini, M., Steidel,
C.C., \& Adelberger, K.L. 2004, ApJ, 612, 108

\bibitem[]{912} Smith, A.M., et al. 2001, ApJ, 546, 829

\bibitem[]{914} Somerville, R.S., Primack, J.R., Faber, S.M. 2001,
MNRAS, 320, 504

\bibitem[]{917} Sommer-Larsen, J., G\"otz, M., \& Portinari, L.
2003, ApJ, 596. 47

\bibitem[]{920} Spergel, D.N., et al. 2003, ApJS, 148, 175

\bibitem[]{922} Steidel, C.C., Giavalisco, M., Dickinson, M., \&
Adelberger, K.L. 1996, AJ, 112, 352

\bibitem[]{925} Taniguchi, Y., \& Shioya, Y. 2001, ApJ, 547, 146

\bibitem[]{927} van den Bergh, S., Abraham, R.G., Ellis, R.S.,
Tanvir, N.R.,Santiago, B.X., \& Glazebrook, K.G. 1996, AJ 112, 359

\bibitem[]{} Vanzella, E., Cristiani, S., Dickinson, M., Kuntschner, H.,
Moustakas, L. A., Nonino, M., Rosati, P., Stern, D., Cesarsky, C.,
Ettori, S., and 7 coauthors, 2004, astro-ph/0406591

\bibitem[]{930} Walker, I., Mihos, J., \& Hernquist, L. 1996, ApJ,
460, 121

\bibitem[]{933} Westera, P., Samland, M., Buser, R., \& Gerhard,
O.E. 2002, A\&A, 389, 761


\end{thebibliography}
\end{document}